\documentclass[conference,10pt, letterpaper]{IEEEtran}
\usepackage{graphicx}
\usepackage{multirow}
\usepackage{url}
\usepackage[ruled, boxed, linesnumbered, commentsnumbered, vlined]{algorithm2e}
\usepackage{subfigure}
\usepackage{amsmath}
\usepackage{amssymb}

\usepackage{color}

\newcommand{\BigO}[1]{\ensuremath{\operatorname{O}\left(#1\right)}}

\everymath{\displaystyle}

\begin{document}

\pdfoutput=1

\title{COSPEDTree-II: Improved Couplet based Phylogenetic Supertree}

\author{\IEEEauthorblockN{Sourya Bhattacharyya, and Jayanta Mukhopadhyay\\}
Department of Computer Science and Engineering, Indian Institute of Technology Kharagpur, \\
West Bengal 721302, India\\
sourya.bhatta@gmail.com, jay@cse.iitkgp.ernet.in}

\maketitle

\begin{abstract}
 A Supertree synthesizes the topologies of a set of phylogenetic trees carrying overlapping 
 taxa set. In process, conflicts in the tree topologies are aimed to be resolved  
 with the consensus clades. Such a problem is proved to be NP-hard. 
 Various heuristics on local search, maximum parsimony, graph cut, etc. lead 
 to different supertree approaches, of which the most popular methods are based 
 on analyzing fixed size subtree topologies (such as triplets or 
 quartets). Time and space complexities of these methods, however, 
 depend on the subtree size considered. Our earlier proposed supertree method 
 COSPEDTree, uses evolutionary relationship among individual couplets (taxa pair), 
 to produce slightly conservative (not fully resolved) supertrees. Here we 
 propose its improved version COSPEDTree-II, which produces better 
 resolved supertree with lower number of missing branches, and incurs much 
 lower running time. Results on biological datasets show that COSPEDTree-II 
 belongs to the category of high performance and computationally 
 efficient supertree methods.
\end{abstract}


\IEEEpeerreviewmaketitle

\section{Introduction}

Supertree methods combine the evolutionary relationships of a set of  
phylogenetic trees \textbf{G}, into a single tree \textbf{T} \cite{BinindaEmonds2014}. These 
methods differ from the consensus-based approaches \cite{Gordon1986, Dong2010}, 
by allowing input trees to have different but overlapping set of taxa. 
Supertrees are useful in combining input trees generated 
from completely incomparable approaches, such 
as statistical analysis of discrete dataset and distance analysis of 
DNA-DNA hybridization data \cite{BinindaEmonds2014}. Input trees often exhibit 
conflicting topologies, due to different evolutionary histories of respective genes, 
stochastic errors in site and taxon sampling, and biological errors due to 
paralogy, incomplete lineage sorting, or horizontal gene transfer \cite{Ranwez2010}. 
Supertree methods quest for resolving such conflicts in order to produce 
a `median tree', which minimizes the sum of a given distance 
measure with respect to the input trees \cite{Ranwez2010}. Large scale supertrees are 
intended towards assembling the \emph{Tree of Life} \cite{Reaz2014}.

Our earlier work \cite{Sourya2014}, and the study in \cite{BinindaEmonds2014}, provide 
a comprehensive review of various supertree methods. \emph{Indirect} supertree methods 
first generate intermediate structures like matrices (as in MRP \cite{Baum1992}, 
Minflip \cite{Chen2006}, SFIT \cite{Creevey2005}) or graphs (as in Mincut (MC) \cite{Semple2000}, 
modified Mincut \cite{Page2002}) from 
the input trees, and subsequently resolve these intermediate structures to 
produce the final supertree. These methods, especially MRP, are quite 
accurate, but computationally intensive. \emph{Direct} methods, 
on the other hand, derives the supertree directly from input tree topologies. 
These methods may aim for minimizing either the sum of false positive (FP) branches 
(as in the \emph{veto} approaches like 
PhySIC \cite{Ranwez2007}, SCM \cite{Roshan2004}) or the sum of 
Robinson-Foulds (RF) \cite{Robinson1981} distance (as in 
RFS \cite{Bansal2010}) between \textbf{T} and \textbf{G}. Another 
approach named Superfine \cite{Swenson2011, Neves2012} employs 
greedy heuristics on MRP \cite{Baum1992} or Quartet Maxcut (QMC) \cite{Snir2010}, 
to derive the supertree, which may not be completely resolved.
Supertrees formed by synthesizing the subtrees (such as triplets \cite{Ranwez2010, Lin2009}, 
quartets \cite{Snir2010, Reaz2014}) of the input trees, exhibit quite high 
performance. But, time and space complexities of these methods 
depend on the size of the subtree used.

We have previously developed COSPEDTree \cite{Sourya2014}, 
a supertree algorithm using evolutionary relationships 
among individual pair of taxa (couplets). The method is computationally 
efficient, but produces somewhat conservative (not fully resolved) 
supertrees, with low number of false positive (FP) but high number of 
false negative (FN) branches between \textbf{T} and \textbf{G}. 
Here we propose its improved version, termed as 
COSPEDTree-II, which produces better resolved supertree, 
with lower number of FN branches between \textbf{T} and \textbf{G}, 
keeping the FP count also low. We have also proposed a mechanism to 
convert a non-resolved supertree into a strict binary tree, to reduce the 
FN count. COSPEDTree-II requires 
significantly lower running time than COSPEDTree and most of the 
reference methods, particularly for the datasets having 
high number of trees or taxa.

Rest of this manuscript is organized as follows. First, 
we review the basics of COSPEDTree (as in \cite{Sourya2014}) 
in section~\ref{sec:overview_cosped}. The method COSPEDTree-II is 
then described in section~\ref{sec:Methodology}. Performance of 
COSPEDTree-II is summarized in section~\ref{sec:Results}.

\section{Overview of COSPEDTree}
\label{sec:overview_cosped}

\begin{figure}[!ht]
\centering
\subfigure[]{\label{fig:source_1}\includegraphics[width=0.4\linewidth,keepaspectratio=true]{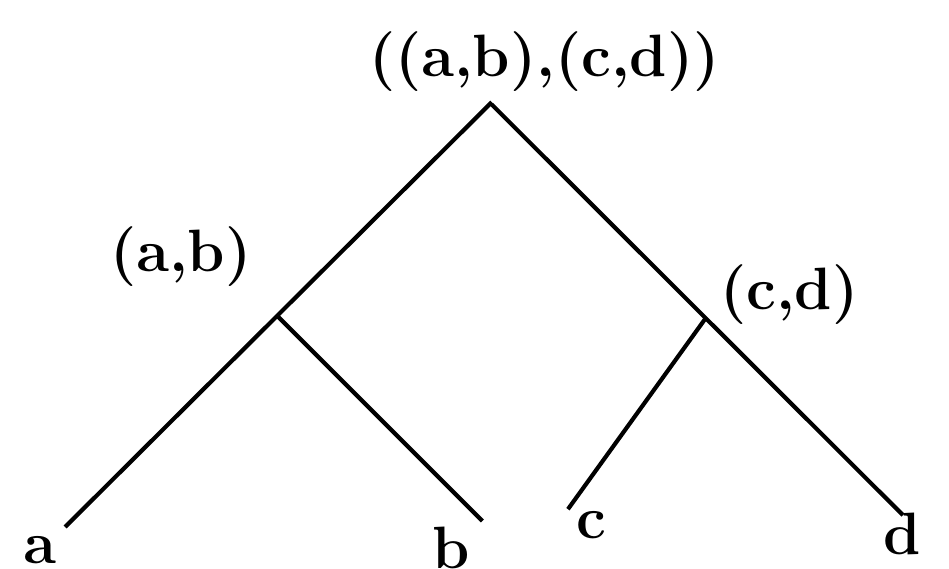}}\qquad
\subfigure[]{\label{fig:source_2}\includegraphics[width=0.4\linewidth,keepaspectratio=true]{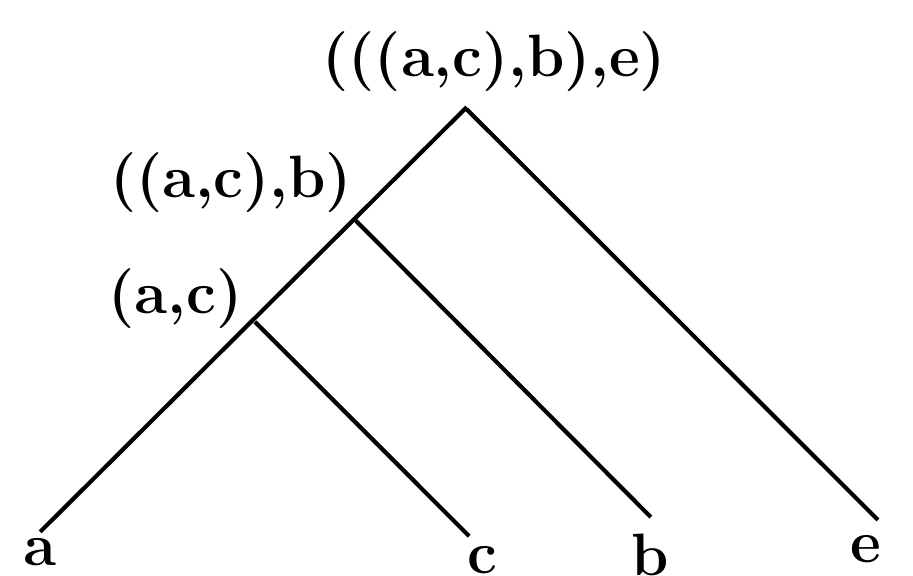}}
\caption{Example input phylogenetic trees. All the nodes are labeled by 
Newick \cite{Sukumaran2010} representation.}
\label{fig:input_trees}
\end{figure}

Let \textbf{G} consist of $M$ rooted input trees $t_1, t_2, \ldots, t_M$. For 
an input tree $t_j$ ($1 \leq j \leq M$), let $L(t_j)$ 
be its set of constituent taxa. Suppose a pair of taxa $p$ and 
$q$ belong to $L(t_j)$. Further, let $\phi_{p}$ and $\phi_{q}$ be the 
parent internal nodes (points of speciation) of $p$ and $q$, respectively. 
COSPEDTree \cite{Sourya2014} defines four 
boolean relations $r^{pq}_{k}$ ($k \in \{ 1,2,3,4 \}$) between $p$ and $q$, 
with respect to $t_j$, as: 

\begin{enumerate}
 \item Earlier Speciation of $p$ than $q$ \textbf{($r^{pq}_{1}$)} is true, 
 if $\phi_{p}$ is ancestor of $\phi_{q}$ in $t_j$. For the tree 
 in Fig.~\ref{fig:source_2}, $r^{bX}_{1}$ is true, where $X \in \{a, c\}$.  
 Similarly, $r^{eY}_{1}$ is true for $Y \in \{a, c, b\}$.
 \item  Later Speciation of $p$ than $q$ \textbf{($r^{pq}_{2}$)} is true, 
 if $\phi_{p}$ is a descendant of $\phi_{q}$. So, $r^{qp}_{1}$ and $r^{pq}_{2}$ 
 are equivalent.
 \item Simultaneous Speciation of $p$ and $q$ \textbf{($r^{pq}_{3}$)} is true, if $\phi_{p}$ = 
 $\phi_{q}$. In Fig.~\ref{fig:source_1}, $r^{ab}_{3}$ and $r^{cd}_{3}$ 
 are true.
 \item Incomparable Speciation of $p$ and $q$ \textbf{($r^{pq}_{4}$)} is true, when 
 $\phi_{p}$ and $\phi_{q}$ occur at different (and independent) 
 clades. For the tree in Fig.~\ref{fig:source_1}, 
 $r^{ac}_{4}$ is true.
\end{enumerate}

Using another taxon $s \in L(t_j)$, properties of $r_{1}$ to $r_{4}$ 
can be stated as the following:

\begin{description}
 \item[P1:] Both $r_{1}$ and $r_{2}$ are transitive. Thus,
 \begin{itemize}
  \item $r^{pq}_{1}$ \& $r^{qs}_{1}$ $\Rightarrow$ $r^{ps}_{1}$.
  \item $r^{pq}_{2}$ \& $r^{qs}_{2}$ $\Rightarrow$ $r^{ps}_{2}$.
 \end{itemize}
 \item[P2:] $r_{3}$ is an equivalence relation.
 \item[P3:] $r^{pq}_{3}$ (= $r^{qp}_{3}$) \& $r^{ps}_{k}$ 
 $\Rightarrow$ $r^{qs}_{k}$, where $k \in \{1, 2, 4\}$. 
 \item[P4:] $r^{pq}_{1}$ (= $r^{qp}_{2}$) \& $r^{ps}_{4}$ (= $r^{sp}_{4}$) 
 $\Rightarrow$ $r^{qs}_{4}$ (= $r^{sq}_{4}$).
\end{description}

\emph{Support tree set} $\varGamma_{pq}$ for a couplet ($p, q$) 
is defined as:
\begin{equation}
\varGamma_{pq} = \{t_j: (p, q) \in L(t_j)\} 
\end{equation}
The \emph{frequency} $F^{pq}_{k}$ ($k \in \{1,2,3,4\}$) of a relation
$r^{pq}_{k}$ between a couplet ($p, q$) is the number of input trees $t_j$ 
where $t_j \in \varGamma_{pq}$ and $r^{pq}_{k}$ is true.

The \emph{set of allowed relations} $R(p,q)$ between a couplet ($p, q$) is 
defined as the following:
\begin{equation}
\label{eq:allowed_reln}
 R(p,q) = \{ r^{pq}_{k}: F^{pq}_{k} > 0 \}
\end{equation}
A couplet ($p, q$) exhibits \emph{conflict} if $|R(p,q)| \geq 2$ 
(where $|.|$ denotes the cardinality of a set). The 
\emph{consensus relation} between ($p, q$) is the relation having 
the maximum frequency.

\textit{Priority} measure $P^{pq}_{k}$ for a relation $r^{pq}_{k}$ \
($k \in \{1,2,3,4\}$) between a couplet ($p, q$) is defined as the following:
\begin{equation}
P^{pq}_{k} = F^{pq}_{k} - \displaystyle \sum_{1 \leq k' \leq 4, k' \neq k} F^{pq}_{k'} 
\end{equation}
COSPEDTree also defines a \emph{support score} $V^{pq}_{k}$ for individual
relations $r^{pq}_{k}$ as the following:
\begin{equation}
V^{pq}_{k} = F^{pq}_{k} \times P^{pq}_{k} 
\end{equation}
The consensus relation $r^{pq}_{k}$ between a couplet ($p, q$) 
exhibits the highest $P^{pq}_{k}$ and $F^{pq}_{k}$ values. So, 
corresponding $V^{pq}_{k}$ also becomes the highest among all 
relations between ($p, q$).

Final supertree \textbf{T} \emph{resolves} (assigns a particular relation to) 
individual couplet ($p, q$) with a single relation $r^{pq}_{k}$ ($k \in \{ 1,2,3,4 \}$) 
between them. Maximum agreement property \cite{Bansal2010} of a
supertree quests for resolving individual couplets with their respective 
consensus relations. But, satisfying such property is NP-hard 
since consensus relations among couplets can be mutually 
conflicting \cite{Sourya2014}. Thus, order of selection of 
individual candidate relations $r^{pq}_{k}$ (to resolve the corresponding 
couplet ($p, q$)) is crucial. In view of this, COSPEDTree first 
constructs a set of relations $S_r$, such that if a relation 
$r^{pq}_{k} \in S_r$, the couplet ($p, q$) is resolved with $r^{pq}_{k}$. 
To construct $S_r$, COSPEDTree applies an iterative greedy approach. 
At each iteration, it selects a relation $r^{p'q'}_{k}$ 
to resolve ($p', q'$) among all unresolved couplets, provided:
\begin{center}
$V^{p'q'}_{k'}$ = $\max_{\substack{\forall (p, q), \forall k}} V^{pq}_{k}$.  
\end{center}
If the selected relation $r^{p'q'}_{k'}$ does not contradict with any of the already 
selected relations in $S_r$ (according to the properties P1 to P4 mentioned 
before), it is included in $S_r$.

\begin{figure}[!ht]
\centering
\subfigure[]{\label{fig:mult_par_1}\includegraphics[width=0.23\linewidth,keepaspectratio=true]{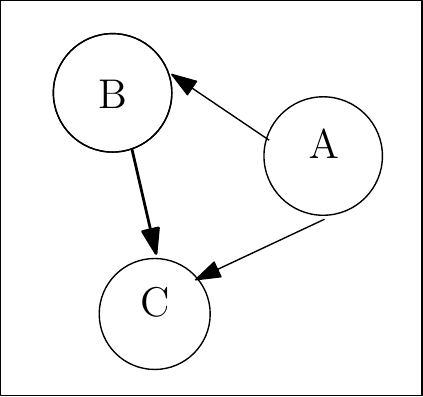}}
\subfigure[]{\label{fig:mult_par_2}\includegraphics[width=0.23\linewidth,keepaspectratio=true]{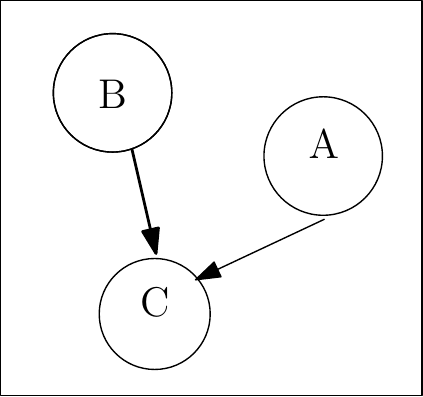}}
\subfigure[]{\label{fig:disc_node}\includegraphics[width=0.23\linewidth,keepaspectratio=true]{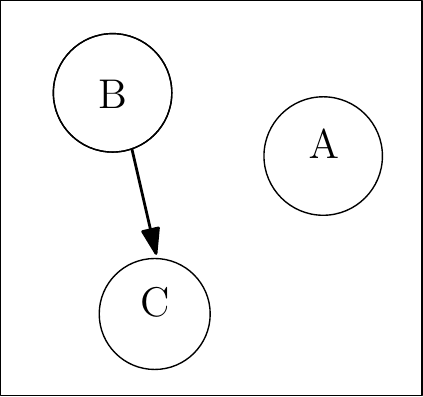}}
\subfigure[]{\label{fig:leaf_reln_dff}\includegraphics[width=0.23\linewidth,keepaspectratio=true]{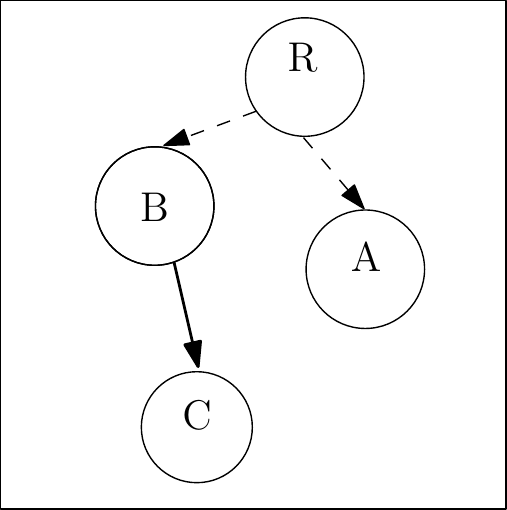}}
\caption{(a) Transitive parent problem (b) Multiple parent problem MPP. 
(c) No parent problem NPP. (d) solution of NPP by COSPEDTree, by inserting 
a hypothetical root R.}
\label{fig:forest_prop}
\end{figure}

Suppose, L(\textbf{G}) = $\cup^{M}_{j=1}L(t_j)$ denotes 
the complete set of input taxa. Then, $N$ = $|$L(\textbf{G})$|$. 
Using the set of relations $S_r$, COSPEDTree partitions L(\textbf{G}) 
into $s$ mutually exclusive taxa clusters 
$C_{1}$,$C_{2}$,$\ldots$,$C_{s}$, with the following rule (details are 
provided in \cite{Sourya2014}):

\begin{description}
 \item[R1:] If a pair of taxa $p$ and $q$ belong to the same cluster $C_i$ 
 ($1 \leq i \leq s$), $r^{pq}_3 \in S_r$. 
 \item[R2:] Suppose $C_{i}$ and $C_{j}$ ($1 \leq i,j \leq s$, $i \neq j$) are 
 any two distinct taxa clusters. Then, $\forall p \in  C_{i}$, and $\forall q \in  C_{j}$, 
 $r^{pq}_k \in S_r$, where $k \in \{1, 2, 4\}$. This property is denoted by 
 saying that $r^{C_iC_j}_k$ is true, or $C_{i}$ is related with $C_{j}$ via the 
 relation $r_k$.
\end{description}

COSPEDTree creates a directed acyclic graph (DAG), whose nodes 
are individual taxa clusters $C_{i}$ ($1 \leq i \leq s$). 
A directed edge from $C_{i}$ to $C_{j}$ means $r^{C_iC_j}_1$ 
is true. However, occurrence of one or more of the following properties means this DAG 
needs to be refined to form a tree:

\begin{enumerate}
 \item Transitive parent problem (TPP): for three nodes A, B, and C, 
 when $r^{AC}_1$, $r^{BC}_1$, and $r^{AB}_1$ are simultaneously true, 
 as indicated in Fig.~\ref{fig:mult_par_1}.
 \item Multiple parent problem (MPP): when $r^{AC}_1$, $r^{BC}_1$, 
 and $r^{AB}_4$ are simultaneously true, as shown in Fig.~\ref{fig:mult_par_2}.
 \item No parent problem (NPP) (Fig.~\ref{fig:disc_node}): when a node $A$ 
 does not have any parent, i.e. So, there exists no node $B$ such that 
 $r^{BA}_1$ is true.
\end{enumerate}

COSPEDTree \cite{Sourya2014} applies transitive reduction to resolve 
TPP. The problem MPP is solved by arbitrary parent 
assignment, while NPP is resolved by assigning one hypothetical 
root node to the isolated node (as shown in Fig.~\ref{fig:leaf_reln_dff}). 
Finally, a depth first traversal of this DAG produces the 
supertree \textbf{T}. As there is no restriction regarding 
the number of taxa in individual taxa clusters 
(partitions with respect to the relation $r_3$), \textbf{T} may not 
be strictly binary (completely resolved).

\section{Proposed Methodology}
\label{sec:Methodology}

COSPEDTree-II extends COSPEDTree by incorporating the following 
modifications:
 
1) COSPEDTree-II skips the formation of $S_r$. Rather, the taxa clusters 
(containing one or more taxon) are first derived, solely by the 
frequencies of different relations between individual couplets. 
Subsequently, directed edges between individual pairs of clusters 
are assigned, according to the properties of individual couplets 
contained within these cluster pairs. Such processing on the taxa clusters, 
rather than the couplets, achieves high speedup and much 
lower running time.
 
2) In COSPEDTree, if a relation $r^{pq}_{k}$ ($1 \leq k \leq 4$) 
between a couplet ($p, q$) is supported in a tree 
$t_j \in \varGamma_{pq}$, the frequency $F^{pq}_{k}$ is 
incremented by 1. COSPEDTree-II, on the other hand, uses 
fractional and dynamic frequency values. In the above case, 
COSPEDTree-II increments $F^{pq}_{k}$ with a \emph{weight} 
$W^{t_j}_{pq}$ ($0 < W^{t_j}_{pq} \leq 1$), which 
varies for individual couplets ($p, q$), and also 
for individual trees $t_j \in \varGamma_{pq}$. 
 
3) For the problem MPP, COSPEDTree-II proposes a 
deterministic selection of the parent, for the 
node having multiple parents.
 
4) COSPEDTree-II also suggests a mechanism to convert a non-binary 
supertree into a binary tree.
 
Subsequent sections describe all such improvements.

\subsection{Fractional frequency value for relations}
\label{subsec:weight_freq}

\begin{figure}[!ht]
\centering
\subfigure[]{\label{fig:no_edge_1}\includegraphics[width=0.23\linewidth,keepaspectratio=true]{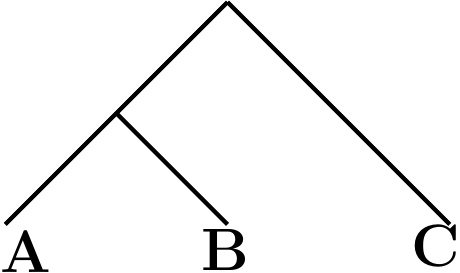}}
\subfigure[]{\label{fig:no_edge_2}\includegraphics[width=0.23\linewidth,keepaspectratio=true]{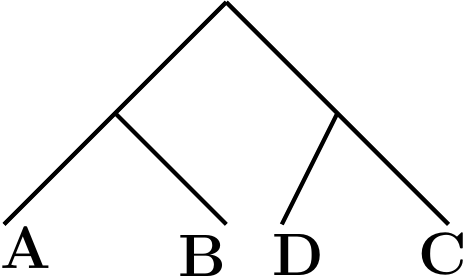}}
\subfigure[]{\label{fig:bidir_edge_1}\includegraphics[width=0.23\linewidth,keepaspectratio=true]{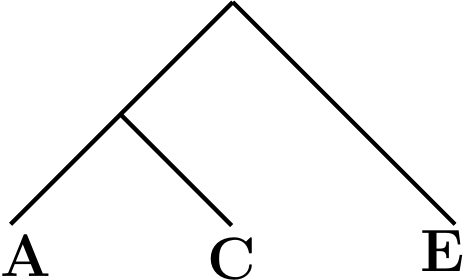}}
\subfigure[]{\label{fig:ex_suptree}\includegraphics[width=0.23\linewidth,keepaspectratio=true]{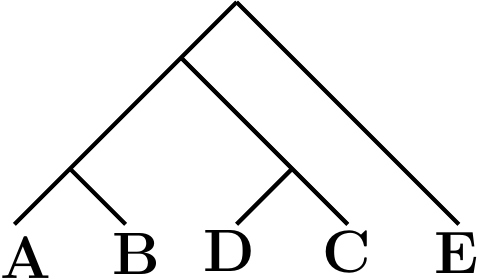}}
\caption{Fig. (a) to (c) shows three input trees. Fig. (d) shows 
the corresponding supertree.}
\label{fig:reln_priority}
\end{figure}

COSPEDTree-II applies a fractional frequency value $W^{t_j}_{pq}$ 
if an input tree $t_j$ supports the relation $r^{pq}_{k}$ 
between a couplet ($p,q$). Value of $W^{t_j}_{pq}$ 
depends on the set $L(t_j)$. Utility of such a dynamic (and fractional) frequency measure 
is explained by Fig.~\ref{fig:reln_priority}, which shows 
three input trees (Fig.~\ref{fig:no_edge_1} to 
Fig.~\ref{fig:bidir_edge_1}) and corresponding supertree (Fig.~\ref{fig:ex_suptree}). 
For the couplet (A,C), all of the relations $r_2$, $r_3$ and $r_4$ are 
supported. However, we observe that the relation $r_3$ is supported only because 
corresponding tree does not include taxa B and D. Similarly, the relation 
$r_2$ occurs due to the absence of the taxon D. When both B and D are 
present (Fig.~\ref{fig:bidir_edge_1}), the relation $r_4$ (which is the ideal 
relation between (A,C)) is satisfied. So, the relation $r_4$ should be 
given higher weight, since the corresponding tree has higher 
taxa coverage. So, our proposed dynamic frequency measure varies according 
to the coverage of taxa of different input trees.

Considering an input tree $t_j$ ($1 \leq j \leq M$) and a couplet ($p,q$) in 
$L(t_j)$, first we define the following notations:

\begin{itemize}
 \item $V(t_j)$: set of nodes (leaf or internal) of $t_j$.
 \item $LCA^{t_j}_{pq}$: lowest common ancestor (LCA) of 
 $p$ and $q$ in $t_j$.
 \item $Clade_{t_j}(v)$: subtree rooted at an internal node 
 $v \in (V(t_j) - L(t_j))$.
 \item $Cluster_{t_j}(v)$: Set of taxa underlying $Clade_{t_j}(v)$.
\end{itemize}

With such definitions, the set of \emph{excess taxa} (excluding the couplet itself) 
underlying the LCA node of ($p,q$) in $t_j$, is defined as the following:
\begin{equation}
 U^{t_j}_{pq} = Cluster_{t_j}(LCA^{t_j}_{pq}) - \{p, q\}
\end{equation}
For ($p,q$), \emph{union of all excess taxa} underlying 
the respective $LCA^{t_j}_{pq}$ nodes for all 
$t_j \in \varGamma_{pq}$, is:
\begin{equation}
 U^{\mbox{\textbf{G}}}_{pq} = \bigcup_{t_j \in \varGamma_{pq}} U^{t_j}_{pq}
\end{equation}
We assign the weight of a relation $r^{pq}_{k}$  
($1 \leq k \leq 4$) between ($p,q$) in an input tree $t_j$, as:
\begin{equation}
\label{eq:weight_couplet_reln}
 W^{t_j}_{pq} = \frac{|U^{\mbox{\textbf{G}}}_{pq} \bigcap L(t_j)|}{|U^{\mbox{\textbf{G}}}_{pq}|}
\end{equation}
where $W^{t_j}_{pq}$ = 1 if $U^{\mbox{\textbf{G}}}_{pq}$ = $\phi$.

Thus, the weight equals the proportion of taxa within 
$U^{\mbox{\textbf{G}}}_{pq}$, that is covered in the input tree 
$t_j$. Frequency $F^{pq}_{k}$ of the relation $r^{pq}_{k}$, is now redefined 
as the following:
\begin{equation}
 F^{pq}_{k} = \sum_{t_j \mbox{ supports } r^{pq}_{k}} W^{t_j}_{pq}
\end{equation}

\subsection{Generating taxa clusters}
\label{subsec:taxa_clust_gen}

COSPEDTree \cite{Sourya2014} creates taxa clusters after formation of the \emph{set 
of resolving relations} $S_r$. COSPEDTree-II, on the other hand, creates taxa 
clusters before resolving any couplets at all. Rather, for individual couplets ($p,q$), 
COSPEDTree-II inspects the values of $F^{pq}_{k}$ for individual relations $r^{pq}_{k}$ 
($k \in \{1, 2, 3, 4\}$). Creation of taxa clusters requires identifying couplets 
which can be resolved by the relation $r_3$. COSPEDTree-II 
places a pair of taxa $p$ and $q$ in the same taxa cluster (thereby 
resolving the couplet ($p,q$) with the relation $r_3$), provided:

\begin{enumerate}
 \item Either $|R(p,q)|$ = 1 and $r^{pq}_{3} \in R(p,q)$ ($R(p,q)$ 
 is already defined in Eq.~\ref{eq:allowed_reln}).
 \item Or $|R(p,q)|$ = 2 and $r^{pq}_{3}$ is majority consensus. In such a case, 
 $F^{pq}_{3} \geq 0.5 * \{\sum_{k} F^{pq}_{k}\}$.
 \item If $|R(p,q)| >$ 2, the couplet ($p,q$) is not placed in the same taxa cluster, 
 even if $r^{pq}_{3}$ is majority consensus. This is because, as the 
 couplet exhibits high degree of conflict, we check the relations between 
 $p$, $q$, and other taxa set.
\end{enumerate}

The first condition is obvious. A couplet having only $r_3$ as its allowed 
relation would be preferably resolved with it. On the other hand, if there 
exists one more relation $r^{pq}_{k'}$ ($k' \neq 3$) within $R(p,q)$, we check whether 
$F^{pq}_{3} > F^{pq}_{k'}$, which ensures that $r^{pq}_{3}$ is 
the majority consensus relation of ($p,q$). In such a case, the 
couplet is highly probable of being resolved with $r_3$ in the final 
supertree.


Above mentioned heuristics are applied for individual couplets, to perform 
the equivalence partitioning (taxa clusters) of the input taxa 
set L(\textbf{G}). 

\subsection{Connectivity between taxa clusters to form DAG}
\label{subsec:taxa_clust_conn}

Creation of the taxa clusters is followed by the assignment of 
directed edges between them. As mentioned in 
section~\ref{sec:overview_cosped}, directed edge from a  
cluster $C_i$ to a cluster $C_j$ corresponds to the 
relation $r^{C_iC_j}_1$ (= $r^{C_jC_i}_2$) 
being true. In such a case, the cluster pair ($C_i$, $C_j$) 
is said to be \emph{resolved by the relation} $r_1$. In general, 
a pair of clusters can be resolved via any one of the relations $r_1$, $r_2$ or 
$r_4$ (no directed edge in this case). 
For individual relations $r_k$ ($k \in \{1, 2, 4\}$), we 
define its frequency $F^{C_iC_j}_k$ with respect to the pair of 
cluster ($C_i, C_j$), as the following:
\begin{equation}
 F^{C_iC_j}_k = \sum_{\forall p \in C_i, \forall q \in C_j} F^{pq}_{k}
\end{equation}
Priority of individual relations $r_k$ ($k \in \{1, 2, 4\}$) 
for the cluster pair ($C_i, C_j$) is defined as the following:
\begin{equation}
 P^{C_iC_j}_{k} = F^{C_iC_j}_{k} - \sum_{k' \in \{1, 2, 4\}, k \neq k'} F^{C_iC_j}_{k'}
\end{equation}
Support score of a relation $r_k$ between the cluster pair ($C_i, C_j$) is 
defined as:
\begin{equation}
 V^{C_iC_j}_{k} = P^{C_iC_j}_{k} + F^{C_iC_j}_{k}
\end{equation}
Note that we have used sum, rather than the product, of the priority and 
frequency measures. This is due to the disparity of signs of frequency (which 
is always non-negative) and the priority (which can be negative even for a 
consensus relation) measures. 
Higher support score of a relation (between a pair of clusters) 
indicates higher frequency and priority of the corresponding relation.

The set $Q$ of support scores for different relations between 
individual cluster pairs is defined as follows:
\begin{equation}
 Q = \{V^{C_iC_j}_{k}: C_i \neq C_j, k \in \{1, 2, 4\}, F^{C_iC_j}_k > 0 \}
\end{equation}
Individual taxa clusters are now resolved by an iterative algorithm, 
using the set $Q$. Each iteration extracts a relation 
$r^{C_xC_y}_{k'}$ ($k' \in \{1, 2, 4\}$)) from $Q$, provided the following:
\begin{equation}
 V^{C_xC_y}_{k'} = \max_{\substack{\forall (C_i, C_j), \forall k}} V^{C_iC_j}_{k}
\end{equation}
Following conditions are checked to see whether the extracted relation $r_{k'}$ 
can resolve the cluster pair ($C_x$, $C_y$). 

\begin{enumerate}
 \item If ($C_x$, $C_y$) is already resolved with a different relation, 
 $r_{k'}$ is not applied. 
 \item If $k'$ = 1 or 2, resolving ($C_x$, $C_y$) with $r_{k'}$ would 
 create a directed edge between the cluster pair. If such an 
 edge forms a cycle with the existing configuration of the taxa clusters, 
 $r_{k'}$ is not applied.
\end{enumerate}

For no such above mentioned conflicts, the relation $r_{k'}$ is applied 
between $C_x$ and $C_y$.

The set $Q$ is implemented as a max-priority queue \cite{Cormen2001}, 
to achieve \BigO{1} time complexity for extracting the cluster 
pair having the maximum support score. Iterations continue until $Q$ becomes 
empty. However, the final DAG may still have the 
problems TPP, MPP, and NPP (as defined in Fig.~\ref{fig:forest_prop}). 
The problem TPP is removed by transitive reduction (already described in 
COSPEDTree \cite{Sourya2014}). COSPEDTree-II employs a 
better solution for the problem MPP, which is 
described in the following section.

\subsection{Solving Multiple Parent Problem (MPP)}
\label{subsec:solnC2}

As shown in Fig.~\ref{fig:mult_par_2}, the problem MPP corresponds to a 
cluster $C_z$ having $k$ ($k \geq$ 2) other clusters $C_1, C_2, \ldots, C_k$ 
as its parent, which are not themselves connected by any directed edges.
The objective is to assign a unique parent $C_p$ ($1 \leq p \leq k$) to the 
cluster $C_z$. Such assignment was arbitrary in COSPEDTree \cite{Sourya2014}. 
COSPEDTree-II proposes a deterministic selection of $C_p$, by a measure 
called the \emph{internode count} $I_{t_j}(p,q)$ between a couplet ($p,q$), 
with respect to a rooted tree $t_j$. The measure was introduced in \cite{Liu2011} 
for unrooted trees. Here, the measure is adapted for a rooted tree $t_j$, 
as the number of internal nodes between $p$ and $q$ 
through the node $LCA^{t_j}_{pq}$.

As individual trees $t_j$ carry overlapping taxa subsets of L(\textbf{G}), 
we define a \emph{normalized internode count distance} 
between $p$ and $q$ in $t_j$ as:
\begin{equation}
I^{N}_{t_j}(p,q) = \frac{I_{t_j}(p,q)}{W^{t_j}_{pq}} 
\end{equation}
where $W^{t_j}_{pq}$ is defined in the Eq.~\ref{eq:weight_couplet_reln}.
So, $I^{N}_{t_j}(p,q)$ becomes low only when both $I_{t_j}(p,q)$ is low 
and $W^{t_j}_{pq}$ is high (when the tree $t_j$ carries higher proportion of 
the taxa subset belonging to $U^{\mbox{\textbf{G}}}_{pq}$).

Significance of the internode count distance can be explained by considering 
a rooted triplet $(r,(p,q))$ (shown in the Newick \cite{Sukumaran2010} format), 
consisting of three taxa $p$, $q$ and $r$. Here, $I^{N}(p,q) < I^{N}(p,r) = I^{N}(q,r)$. 
In general, lower internode count means corresponding 
couplet is evolutionarily closer, compared to the other couplets.

\emph{Average internode count} of a couplet $(p,q)$, with respect 
to \textbf{G}, is defined by the following expression:
\begin{equation}
 I_{avg}(p,q) = \frac{1}{|\varGamma_{pq}|} \sum_{t_j \in \varGamma_{pq}} I^{N}_{t_j}(p,q)
\end{equation}
The \emph{internode count distance between a pair of cluster} $C_x$ and 
$C_y$ is defined by the following equation:
\begin{equation}
 I(C_x, C_y) = \frac{\sum_{\forall p \in C_x, q \in C_y} I_{avg}(p,q)}{|C_x||C_y|} 
\end{equation}
where $|C_x|$ denotes the cardinality of the taxa cluster $C_x$. 

For the MPP problem, COSPEDTree-II selects the cluster $C_p$ ($1 \leq p \leq k$) 
as the parent of $C_z$, provided that $C_p$ has the lowest internode count 
distance to $C_z$:
\begin{equation}
 C_p = \mbox{argmin}_{\substack{1 \leq i \leq k}} I(C_z, C_i)
\end{equation}
Such condition is based on the assumption that the cluster pair 
having lower internode count, is possibly closer in 
the evolutionary tree, compared to other cluster pairs.

\subsection{Binary supertree generation}
\label{subsec:refine_tree}

\begin{figure}[!ht]
\centering
\subfigure[]{\label{fig:multi_furc_clust}\includegraphics[width=0.8\linewidth,keepaspectratio=true]{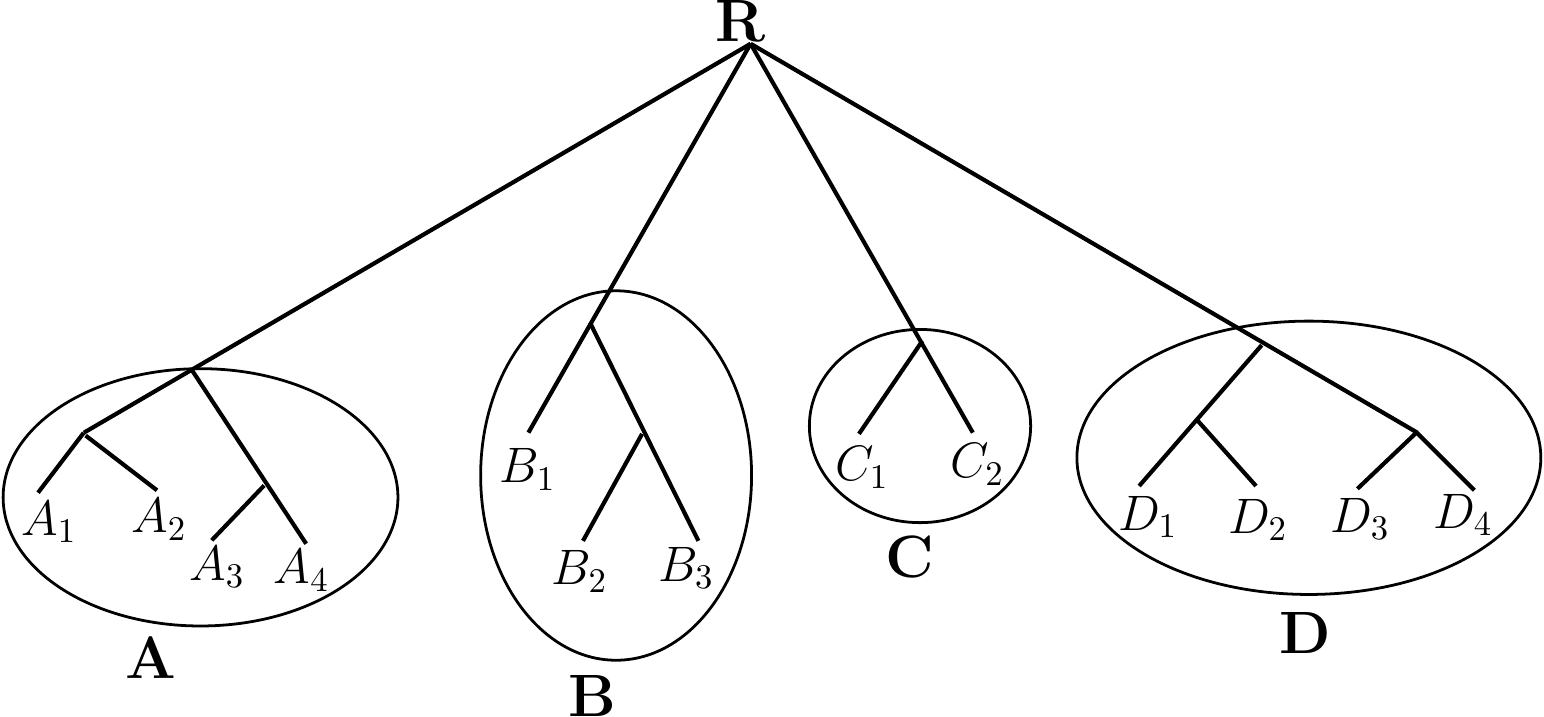}}
\subfigure[]{\label{fig:multi_furc}\includegraphics[width=0.45\linewidth,keepaspectratio=true]{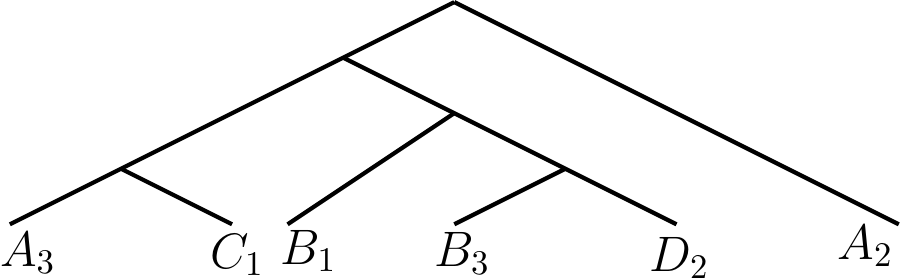}}
\subfigure[]{\label{fig:multi_furc_refine}\includegraphics[width=0.45\linewidth,keepaspectratio=true]{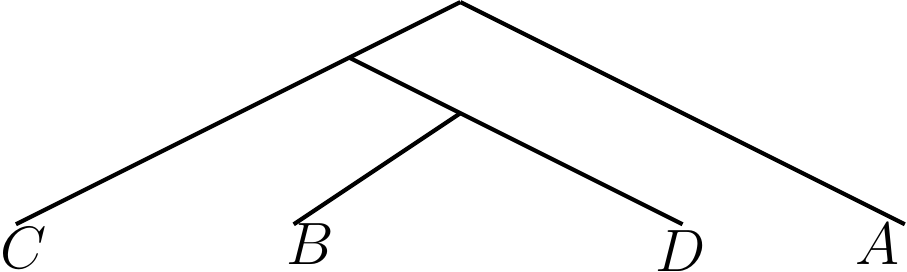}}
\caption{(a) Example of a multifurcation, containing the taxa 
subset $X = A \cup B \cup C \cup D$. (b) an input tree $t_{j|X}$, 
restricted to the taxa subset $X$. (c) Tree $t'_{j|X}$ created from $t_{j|X}$.}
\label{fig:binary_refine}
\end{figure}

After resolving the problem MPP, the refined DAG is converted to the 
supertree \textbf{T}, by a depth first traversal procedure (as 
described in COSPEDTree \cite{Sourya2014}). However, the generated 
supertree \textbf{T} may not be completely resolved. COSPEDTree-II 
proposes a refinement strategy which converts \textbf{T} into a 
strict binary tree.

Suppose, the tree contains an internal multi-furcating node of degree 
$n$ ($> 2$). Let $X_1, X_2, \ldots, X_n$ denote the taxa subsets 
descendant from it, where each taxa subset $X_i$ ($1 \leq i \leq n$) consists 
of one or more taxon named as $X_{i1}, X_{i2}, \ldots$, etc. 
Union of these taxa subsets is represented by $X$ = $\cup^{n}_{i=1} X_{i}$.
Suppose, $t_{j|X}$ represents the input tree $t_j$ ($1 \leq j \leq M$) 
\emph{restricted} to the set of taxa $X$. Thus, $L(t_{j|X})$ = 
$L(t_{j}) \cap X$. Considering Fig.~\ref{fig:multi_furc_clust} as 
an example, the node $R$ represents a multi-furcation with degree 4. 
Four taxa subsets A, B, C, and D, descend from R. Here, 
$X = A \cup B \cup C \cup D$. Generation of a binary 
tree requires introducing bifurcations among these 
taxa subsets. So, for individual input trees $t_j$, 
corresponding restricted input tree $t_{j|X}$ is produced, 
as shown in Fig.~\ref{fig:multi_furc}.



Our proposed binary refinement first generates a tree $t'_{j|X}$ 
from the tree $t_{j|X}$, such that 
the leaves of $t'_{j|X}$ represent individual taxa subsets $X_i$ 
($1 \leq i \leq n$). In other words, individual taxon in $t_{j|X}$ 
is replaced by the name of its corresponding taxa subset 
(without any duplicate). For example, both the taxa $A_2$ and $A_3$ 
(belonging to the taxa subset $A$) are present in the tree $t_{j|X}$ (as shown in 
Fig.~\ref{fig:multi_furc}). So, in $t'_{j|X}$, a leaf node labeled $A$ 
is first inserted as a child of the LCA node of $A_2$ and $A_3$. 
Subsequently, the leaves $A_2$ and $A_3$ are deleted from $t'_{j|X}$. 
This process is repeated for other taxa subsets $B$, $C$ and $D$ 
as well. Fig.~\ref{fig:multi_furc_refine} shows the tree $t'_{j|X}$.

For the current set of taxa $X$, each of the input trees $t_j$ 
are processed to generate the corresponding $t'_{j|X}$. 
These trees are then used as input 
to an existing triplet based supertree approach  
thTBR \cite{Lin2009}, to generate a supertree $T_{X}$ consisting 
of the taxa subsets $X_i$ as its leaves. The supertree method is selected 
since it processes rooted triplets, and generates a 
rooted output tree. The tree $T_{X}$ is used as a template, such that its 
order of bifurcation among individual taxa subsets $X_i$ is 
replicated to the original multi-furcating node $R$ and 
its descendants. As the degree of 
multifurcation ($n$ in this case) is much lower than the total number 
of taxa ($N$), construction of $T_{X}$ is very fast. This process is continued 
until all the multi-furcating nodes are resolved.

%
%

\begin{table}\center\scriptsize
\caption{Results for Marsupials \cite{Cardillo2004} dataset 
($M$ = 158, $N$ = 267)}
\begin{tabular}{|p{2.4cm}|p{0.5cm}|p{0.5cm}|p{0.5cm}|p{0.7cm}|p{1.5cm}|}\hline
 \textbf{Method} & \textbf{FP} & \textbf{FN} & 
 \textbf{RF} & \textbf{MAST} & \textbf{Runtime} \\\hline
 Minflip$^{*}$ \cite{Chen2006} &792 &946 &1738 &75.84 &20m \\
 MMC$^{*}$ \cite{Page2002} &911 &1251 &2162 &69.3 & - \\
 MRP PAUP$^{*}$ \cite{Baum1992} &756 &400 &1156 &86.59 &4.6m  \\
 PhySIC \cite{Ranwez2007} &0 &1324 &1324 &35.1 &14s \\
 RFS \cite{Bansal2010} &710 &361 &1071 &105.6 &6.5m \\
 SCM$^{*}$ \cite{Roshan2004} &0 &1220 &1220 &40.75 & - \\
 SFIT$^{*}$ \cite{Creevey2005} &1327 &979 &2306 &61.69 &111h \\
 Superfine(MRP)$^{*}$ \cite{Swenson2011} &750 &396 &1346 &89.54 &3m  \\
 thSPR \cite{Lin2009} &740 &860 &1600 &97.26 &3m \\
 thTBR \cite{Lin2009} &739 &859 &1598 &97.1 &3m \\
 Supertriplet \cite{Ranwez2010} &598 &390 &988 &ER &57s \\
 COSPEDTree &326 &841 &1167 &68.93 &6m \\
  \textbf{COSPEDTree-II} &\textbf{459} &\textbf{695} &\textbf{1154} &\textbf{78.02} &\textbf{2.4s + 2.2s} \\
  \textbf{COSPEDTree-II+B} &\textbf{827} &\textbf{482} &\textbf{1309} &\textbf{89.34} &\textbf{2.4s+2.2s+34.4s} \\\hline
\end{tabular}
\label{tab:perf_cospedtree_marsupials}
\end{table}

\begin{table}\center\scriptsize
\caption{Results for Placental Mammals \cite{Beck2006} ($M$ = 726, $N$ = 116)}
\begin{tabular}{|p{2.4cm}|p{0.5cm}|p{0.5cm}|p{0.5cm}|p{0.7cm}|p{1.5cm}|}\hline
 \textbf{Method} & \textbf{FP} & \textbf{FN} & 
 \textbf{RF} & \textbf{MAST} & \textbf{Runtime} \\\hline
 Minflip$^{*}$ \cite{Chen2006} &2965 &4002 &6967 &276.64 &7.25m \\
 MRP PAUP$^{*}$ \cite{Baum1992} &2545 &1902 &4447 &313.77 &3.5m  \\
 PhySIC \cite{Ranwez2007} &0 &4830 &4830 &222.86 &3s \\
 RFS \cite{Bansal2010} &2481 &1650 &4131 &511.54 &4.2m \\
 SCM$^{*}$ \cite{Roshan2004} &0 &4816 &4816 &223.86 & - \\
 SFIT$^{*}$ \cite{Creevey2005} &3315 &2353 &5668 &316.53 &108h \\
 QIMP$^{*}$ \cite{Holland2007} & 2423 & 1823 & 4246 & 477.84 & - \\
 Superfine(MRP)$^{*}$ \cite{Swenson2011} &2540 &1746 &4286 &439.65 &9.25m  \\
 Superfine(QMC)$^{*}$ \cite{Swenson2011} &2631 &1835 &4466 &432.84 & 6m \\
 thSPR \cite{Lin2009} &2627 &3268 &5895 &496.23 &7s \\
 thTBR \cite{Lin2009} &2626  &3272  &5898 &496.25 &6.39s \\
 Supertriplet \cite{Ranwez2010} & F & F & F & F & F \\
 COSPEDTree &1232 &3762 &4994 &394.42 &3.2m \\
  \textbf{COSPEDTree-II} &\textbf{1406} &\textbf{2601} &\textbf{4007} &\textbf{429.34} &\textbf{3.5s + 0.5s} \\
  \textbf{COSPEDTree-II+B} &\textbf{2730} &\textbf{1819} &\textbf{4549} &\textbf{491.85} &\textbf{3.5s+0.5s+9.2s} \\\hline
\end{tabular}
\label{tab:perf_cospedtree_placental_mammals}
\end{table}

\begin{table}\center\scriptsize
\caption{Results for Seabirds \cite{Kennedy2002} dataset 
($M$ = 7, $N$ = 121)}
\begin{tabular}{|p{2.4cm}|p{0.5cm}|p{0.5cm}|p{0.5cm}|p{0.7cm}|p{1.5cm}|}\hline
 \textbf{Method} & \textbf{FP} & \textbf{FN} & 
 \textbf{RF} & \textbf{MAST} & \textbf{Runtime} \\\hline
 Minflip$^{*}$ \cite{Chen2006} &38 &72 &110 &3.24 &11s \\
 MRP PAUP$^{*}$ \cite{Baum1992} &61 &166 &227 &4.97 &11s  \\
 PhySIC \cite{Ranwez2007} &0 &150 &150 &0.64 &3s \\
 RFS \cite{Bansal2010} &28 &14 &42 &5.63 &4s \\
 SCM$^{*}$ \cite{Roshan2004} &1 &66 &67 &2.75 & - \\
 SFIT$^{*}$ \cite{Creevey2005} &42 &202 &244 &2.22 &1h \\
 Superfine(MRP)$^{*}$ \cite{Swenson2011} &32 &19 &51 &4.43 &3s  \\
 Superfine(QMC)$^{*}$ \cite{Swenson2011} &29 &19 &48 &4.6 & 3s \\
 thSPR \cite{Lin2009} &69 &242 &311 &5.71 &6s \\
 thTBR \cite{Lin2009} &115 &234 &349 &5.73 &6s \\
 Supertriplet \cite{Ranwez2010} & 2 & 176 & 178 & ER & 5.6s \\
 COSPEDTree &24 &81 &105 &2.37 &3s \\
  \textbf{COSPEDTree-II} &\textbf{27} &\textbf{66} &\textbf{93} &\textbf{2.58} &\textbf{0.2s + 0.3s} \\
  \textbf{COSPEDTree-II+B} &\textbf{50} &\textbf{37} &\textbf{86} &\textbf{5.02} &\textbf{0.2s+0.3s+3.1s} \\\hline
\end{tabular}
\label{tab:perf_cospedtree_seabirds}
\end{table}

\begin{table}\center\scriptsize
\caption{Results for THPL \cite{Wojciechowski2000} dataset
($M$ = 19, $N$ = 558)}
\begin{tabular}{|p{2.4cm}|p{0.5cm}|p{0.5cm}|p{0.5cm}|p{0.7cm}|p{1.5cm}|}\hline
 \textbf{Method} & \textbf{FP} & \textbf{FN} & 
 \textbf{RF} & \textbf{MAST} & \textbf{Runtime} \\\hline
 Minflip$^{*}$ \cite{Chen2006} &142 &149 &291 &4.93 &1.1h \\
 MRP PAUP$^{*}$ \cite{Baum1992} &75 &476 &551 &6.27 &31m   \\
 PhySIC \cite{Ranwez2007} &0 &279 &279 &1.19 &5.7m \\
 RFS \cite{Bansal2010} &106 &66 &172 &11.9 &4.5m \\
 SCM$^{*}$ \cite{Roshan2004} &13 &128 &141 &4.64 & - \\
 Superfine(MRP)$^{*}$ \cite{Swenson2011} &85 & 50 &135 &6.39 &1m  \\
 Superfine(QMC)$^{*}$ \cite{Swenson2011} &62 &43 &105 &6.5 & 1.5m \\
 thSPR, thTBR \cite{Lin2009} & ER & ER & ER & ER & ER  \\
 Supertriplet \cite{Ranwez2010} & F & F & F  &F  &F \\
 COSPEDTree &88 &162 &250 &4.21 &4.5m \\
  \textbf{COSPEDTree-II} &\textbf{96} &\textbf{137} &\textbf{233} &\textbf{5.74} &\textbf{2s + 9s} \\
  \textbf{COSPEDTree-II+B} &\textbf{166} &\textbf{114} &\textbf{280} &\textbf{8.22} &\textbf{2s+9s+1.2m} \\\hline
\end{tabular}
\label{tab:perf_cospedtree_THPL}
\end{table}

\begin{table}\center\scriptsize
\caption{Results for Cetartiodactyla \cite{Price2005} dataset
($M$ = 201, $N$ = 299)}
\begin{tabular}{|p{2.4cm}|p{0.5cm}|p{0.5cm}|p{0.5cm}|p{0.7cm}|p{1.5cm}|}\hline
 \textbf{Method} & \textbf{FP} & \textbf{FN} & 
 \textbf{RF} & \textbf{MAST} & \textbf{Runtime} \\\hline
 MMC$^{*}$ \cite{Page2002} &1181 &1438 &2619 &83.84 & - \\
 MRP PAUP$^{*}$ \cite{Baum1992} &860 &964 &1824 &120.84 & -   \\
 PhySIC \cite{Ranwez2007} &ER &ER &ER &ER &ER \\
 RFS \cite{Bansal2010} &ER &ER &ER &ER &ER \\
 thSPR \cite{Lin2009} &969 &1006 &1975 &118.39 &5.5m  \\
 thTBR \cite{Lin2009} &969 &1006 &1975 &118.09 &4.5m \\
 Supertriplet \cite{Ranwez2010} & 125 & 2175 & 2300 & ER & 59s \\
 COSPEDTree &510 &1001 &1511 &80.43 &11.7m \\
  \textbf{COSPEDTree-II} &\textbf{732} &\textbf{864} &\textbf{1566} &\textbf{95.2} &\textbf{2s + 1s} \\
  \textbf{COSPEDTree-II+B} &\textbf{1240} &\textbf{667} &\textbf{1907} &\textbf{102.18} &\textbf{2s+1s+43s} \\\hline
\end{tabular}
\label{tab:perf_cospedtree_Cetartiodactyla}
\end{table}

\begin{table}\center\scriptsize
\caption{Results for Mammal \cite{Ranwez2007, Ranwez2010} dataset
($M$ = 12958, $N$ = 33)}
\begin{tabular}{|p{2.4cm}|p{0.6cm}|p{0.6cm}|p{0.6cm}|p{0.6cm}|p{1.4cm}|}\hline
 \textbf{Method} & \textbf{FP} & \textbf{FN} & 
 \textbf{RF} & \textbf{MAST} & \textbf{Runtime} \\\hline
 PhySIC \cite{Ranwez2007} &17414 &254527 &271941 &968 &36s \\
 RFS \cite{Bansal2010} &ER &ER &ER &ER &ER \\
 thSPR \cite{Lin2009} & 272752 & 296159 & 568911 & 8378 & 7s  \\
 thTBR \cite{Lin2009} & 276104 & 301787 & 577891 & 8378 & 7s \\
 Supertriplet \cite{Ranwez2010} & 71117 & 105671 & 176788 & ER &6s \\
 COSPEDTree &35124 &141295 &176419 &4441.5 &5.3m \\
 \textbf{COSPEDTree-II} &\textbf{39079} &\textbf{134834} & \textbf{173913} &\textbf{4577.26} &\textbf{2m+0.1s} \\
 \textbf{COSPEDTree-II+B} &\textbf{104226} &\textbf{98407} & \textbf{202633} &\textbf{8365.57} &\textbf{2m+0.1s+1m} \\\hline 
\end{tabular}
\label{tab:perf_cospedtree_Mammal}
\end{table}

\subsection{Computational complexity of COSPEDTree-II}
\label{subsec:complexity}

For $M$ input trees covering a total of $N$ taxa, both COSPEDTree 
\cite{Sourya2014} and COSPEDTree-II incurs \BigO{MN^2} time complexity 
for extracting the couplet based measures from the trees. These 
methods differ in their subsequent steps. 
COSPEDTree first resolves individual couplets in \BigO{N^2\lg{N}} time 
(as shown in \cite{Sourya2014}), and subsequently partitions the taxa 
set according to the relation $r_3$, to form a DAG containing 
$N_C$ ($< N$) nodes (taxa clusters). Formation of a supertree from 
this DAG involves \BigO{N_C^3} time complexity \cite{Sourya2014}.

COSPEDTree-II, on the other hand, first forms 
the taxa clusters in \BigO{N^2} time (processing time for all 
couplets). Subsequently, support scores for individual relations between each pair of 
taxa clusters are placed in the max-priority queue $Q$. 
Here, size of $Q$ is \BigO{N_C^2}, considering $N_C$ as the 
number of taxa clusters. During each iteration, maintaining the 
max-priority property of $Q$ requires \BigO{\lg{N_C}} time. So, 
the complete iterative stage to resolve all pairs of clusters 
(assigning connectivities between them) involves 
\BigO{N_C^2\lg{N_C}} time complexity. As in general, 
$N_C$ is considerably lower than $N$, this iterative step in 
COSPEDTree-II is much faster than COSPEDTree.

Resolving individual pair of clusters, rather than the couplets, 
enables COSPEDTree-II to achieve a significant speedup. 
Suppose, $|X|$ denotes the cardinality of a taxa cluster $X$. So, 
for a pair of taxa clusters $X$ and $Y$, COSPEDTree \cite{Sourya2014} 
resolves all $|X|$ $\times$ $|Y|$ couplets, and maintains their 
relations (and the transitive connectivities inferred 
from these relations). But COSPEDTree-II resolves $X$ and $Y$ 
by processing only one relation between them. So, for this 
cluster pair, speedup achieved by COSPEDTree-II is 
$\approx$ $|X|$ $\times$ $|Y|$. 
For a total of $N_C$ taxa clusters, number of cluster pairs 
is $\binom{N_C}{2}$. Thus, overall speedup achieved by 
COSPEDTree-II is $\approx$ 
$\sum_{X, Y \in \binom{N_C}{2}}$ $|X|$ $\times$ $|Y|$.

To derive the time complexity associated with the binary 
refinement of COSPEDTree-II, suppose $m$ is the number 
of internal nodes in \textbf{T} having degree $> 2$. 
Further, suppose $n$ ($> 2$) denotes the maximum degree of 
multi-furcation among all of these nodes. In such a case, 
applying thTBR \cite{Lin2009} for a particular internal 
node involves maximum \BigO{Mn^3} time complexity. So, overall 
complexity of the binary refinement stage is \BigO{Mn^3m}.

COSPEDTree \cite{Sourya2014} involves a storage complexity 
of \BigO{N^2}, to store the couplet based measures. 
COSPEDTree-II uses additional storage space for storing 
the set of excess taxa $U^{\mbox{\textbf{G}}}_{pq}$ for 
individual couplets ($p,q$). As $0 \leq |U^{\mbox{\textbf{G}}}_{pq}| \leq (N - 2)$, 
the space complexity of COSPEDTree-II is \BigO{N^3}.

\section{Experimental Results}
\label{sec:Results}

Both COSPEDTree and COSPEDTree-II are implemented in Python (version 2.7). 
Tree topologies are processed by the 
phylogenetic library Dendropy \cite{Sukumaran2010}. A desktop 
having Intel\textsuperscript{\textregistered} Quad CoreIntel\textsuperscript{\texttrademark}
i5-3470 CPU, with 3.2 GHz processor and 8 GB RAM, is used to 
execute these methods.

\subsection{Dataset}

COSPEDTree-II is tested with the datasets like 
Marsupials (267 taxa and 158 input trees) \cite{Cardillo2004}, 
Placental Mammals (726 trees and 116 taxa) \cite{Beck2006}, 
Seabirds (121 taxa and 7 trees) \cite{Kennedy2002}, 
Temperate Herbaceous Papilionoid Legumes (THPL) (19 trees and 
558 taxa) \cite{Wojciechowski2000}. Work in \cite{Swenson2011} 
modified these datasets by removing duplicate taxon names and 
few infrequent taxa information\footnote{Datasets are downloaded 
from the link \url{http://www.cs.utexas.edu/~phylo/software/superfine/submission/.}}. 
We have also experimented with Mammal dataset \cite{Ranwez2007, Ranwez2010} 
consisting of 12958 trees and 33 taxa\footnote{Downloaded from the link 
http://www.supertriplets.univ-montp2.fr/.}. In addition, the dataset 
Cetartiodactyla (201 input trees and 299 taxa) \cite{Price2005} is also 
tested\footnote{Maintained in TreeBASE \cite{Sanderson1994}, 
and is downloaded from the link 
\url{https://treebase.org/treebase-web/search/study/summary.html?id=1271}.}.

\subsection{Performance measures}

Performance comparison between COSPEDTree-II and the reference approaches, 
employs the following measures:

\begin{enumerate}
 \item \textbf{False positive distance} FP(\textbf{T}, $t_j$): Number of internal branches 
 present in the supertree \textbf{T}, but not in the input tree $t_{j}$ 
 $(1 \leq j \leq M)$.
 \item \textbf{False negative distance} FN(\textbf{T}, $t_j$): Number of internal branches 
 present in $t_{j}$ but not in \textbf{T}.
 \item \textbf{Robinson-Foulds distance} RF(\textbf{T}, $t_j$): 
 Defined as FP(\textbf{T}, $t_j$) + FN(\textbf{T}, $t_j$).
 \item \textbf{Maximum agreement subtree} MAST(\textbf{T}, $t_j$):  
 Let $N_1$ be the number of taxa contained in the maximum 
 agreement subtree (MAST) common to $T$ and $t_{j}$. Then, 
 MAST(\textbf{T}, $t_j$) = $\frac{N_1}{|L(t_j)|}$. This measure is 
 computed using Phylonet \cite{Than2008}.
\end{enumerate}

Above measures are accumulated for all 
the input trees $t_{j}$ $(1 \leq j \leq M)$, to be used as the final 
performance measures. Supertree producing lower values of 
the sum of FP, FN, and RF values is considered better. On the other hand, 
supertree having higher sum of MAST score is considered superior.

\subsection{Performance comparison}

We have reported the results for the following two variations of COSPEDTree-II:

\begin{enumerate}
 \item \textbf{COSPEDTree-II}: Produces supertree with possible multi-furcations.
 \item \textbf{COSPEDTree-II + B}: Produces completely resolved binary supertrees, 
 by applying the binary refinement suggested in section~\ref{subsec:refine_tree}.
\end{enumerate}

Tables~\ref{tab:perf_cospedtree_marsupials} 
to \ref{tab:perf_cospedtree_Mammal} compare the performances  
of both of these variants, and with the reference approaches as well. 
Reference methods marked with a symbol `*', could 
not be executed in all datasets, either due to the unavailability of 
corresponding source code, or due to their very high computational complexity.
In such a case, we have used their results (both 
topological performance and running time) published in 
the existing studies \cite{Swenson2011}. The approaches MRP and 
superfine require PAUP* \cite{Swofford2003} to execute, which is a
commercial tool and not available to us. So, these methods could not 
be tested in all datasets. Missing entries are indicated by `-'. 

The methods RFS \cite{Bansal2010} and 
Supertriplet \cite{Ranwez2010} produced errors in parsing 
few of the input datasets. Entries showing `ER' 
indicate these errors. Supertrees generated by 
Supertriplet \cite{Ranwez2010} could not be 
parsed by Phylonet \cite{Than2008}. So, we could not compute the 
MAST scores for these trees. Finally, 
a symbol `F' indicates that corresponding method could not 
produce a valid supertree for that dataset.

Results show that COSPEDTree-II produces better resolved 
supertrees than COSPEDTree, as indicated by 
lower FN, and mostly lower RF values for individual datasets. 
COSPEDTree-II also achieves higher MAST scores for these datasets. 
COSPEDTree-II+B produces completely resolved 
binary supertrees. So, the number of FN branches reduces. 
However, as the input trees 
may not be fully resolved (may contain multi-furcating nodes), 
number of FP branches increases considerably. 
As COSPEDTree-II+B produces completely resolved 
supertrees, corresponding MAST scores are much higher than COSPEDTree-II.

Comparison with reference approaches shows that only 
RFS \cite{Bansal2010} produces supertrees with consistently lower RF 
and higher MAST scores than COSPEDTree-II. The method Superfine \cite{Swenson2011} 
performs better than COSPEDTree-II for the datasets Seabirds and 
THPL, while our methods perform slightly better (in terms of lower RF and 
higher MAST score) for the Marsupials and Placental Mammals dataset. 
Superfine does not always generate strictly binary (completely resolved) 
supertrees (for example, in the THPL dataset), unlike COSPEDTree-II+B. 
Such a supertree exhibits much lower RF, but also much lower 
MAST score (compared to COSPEDTree-II+B).

Matrix based methods like Minflip, SFIT, MMC, are outperformed by 
COSPEDTree-II. Veto approaches like SCM, PhySIC, produce supertrees with 
the lowest (mostly zero) FP branches, by not including any conflicting 
clades. In such a case, the number of FN branches becomes very high, 
and MAST scores of these trees also become much lower. COSPEDTree-II also 
produces significantly better results than MRP PAUP for all the datasets 
except Cetartiodactyla. Subtree decomposition based approaches like thSPR, thTBR, 
produce slightly higher MAST score values than COSPEDTree-II, since these 
methods directly synthesize input triplets, or in general, subtree topologies. 
Considering the measure RF, on the other hand, these methods are mostly outperformed 
by COSPEDTree-II.

\subsection{Runtime Comparison}

Tables~\ref{tab:perf_cospedtree_marsupials} to \ref{tab:perf_cospedtree_Mammal} 
express the running time of COSPEDTree-II and COSPEDTree-II+B for different 
datasets, in the formats (A+B) or (A+B+C), respectively, where:

\begin{enumerate}
 \item A = Time to extract the couplet based measures from the input trees.
 \item B = Time to process the couplets and cluster pairs, to produce 
 a (possibly not binary) supertree.
 \item C = Time required to refine the non-resolved supertree into a 
 strict binary tree.
\end{enumerate}

We observe that COSPEDTree-II incurs a significant fraction of its 
running time in the stages A and C. The stage A depends on 
the processing speed of the python based phylogenetic library Dendropy 
\cite{Sukumaran2010}. On the other hand, 
running time for the stage C depends both 
on the construction of $t'_{j|X}$ from individual $t_j$ 
for all the multi-furcating nodes, and on the execution of 
thTBR \cite{Lin2009}. Results show that COSPEDTree-II 
incurs much lower running time than COSPEDTree. 
Excluding the binary refinement stage, the running time is decreased by 
a factor from 2 (for the dataset Mammal \cite{Ranwez2007, Ranwez2010}) to 135 
(for the dataset Cetartiodactyla \cite{Price2005}).

When the number of taxa is high (such as Marsupials \cite{Cardillo2004}, 
Cetartiodactyla \cite{Price2005}), COSPEDTree-II exhibits much 
lower running time than the triplet based methods \cite{Lin2009}, 
\cite{Ranwez2010}, due to its lower time complexity. For 
datasets with large number of trees, COSPEDTree-II incurs 
slightly higher running time than these methods, due to 
the time associated in extracting the couplet based measures.

\section{Conclusion}

We have proposed COSPEDTree-II, an improved couplet based supertree construction 
method (extending our earlier proposed method COSPEDTree). 
COSPEDTree-II produces supertrees with lower topological errors, 
and incurs much lower running time (compared to COSPEDTree). A 
binary refinement to generate a fully resolved supertree, is also 
suggested. Due to its high performance and much lower running time, 
COSPEDTree-II can be applied in large scale biological datasets.

\section*{Executable}

Executable and the results of COSPEDTree-II are provided in the link 
\url{http://www.facweb.iitkgp.ernet.in/~jay/phtree/cospedtree2/cospedtree2.html}.

\section*{acknowledgements}

The first author acknowledges Tata Consultancy Services (TCS) for providing the research 
scholarship. 

\bibliographystyle{ieeetr}

\end{document}